\pdfoutput=1
\documentclass[letterpaper,twocolumn,fp]{jpsj3}
\pdfpagewidth=\paperwidth
\pdfpageheight=\paperheight
\usepackage{graphicx} 
\usepackage{bm} 

\begin{document}
\title{Elastocaloric Response of PbTiO$_3$ Predicted from a First-Principles Effective Hamiltonian}
\author{Jordan A. Barr$^{1}$, Scott P. Beckman$^{1}$, and Takeshi Nishimatsu$^{2}$\thanks{E-mail: t-nissie@imr.tohoku.ac.jp}}
\inst{$^{1}$Department of Materials Science and Engineering, Iowa State University, Ames, Iowa 50011, USA\\$^{2}$Institute for Materials Research (IMR), Tohoku University, Sendai 980-8577, Japan}

\abst{
  A first-principles effective Hamiltonian is used
  in a molecular dynamics simulation to study the elastocaloric effect
  in PbTiO$_3$.
  It is found that the transition temperature is a linear function of uniaxial tensile stress.
  A negative temperature change is calculated, when the uniaxial tensile stress is switched off,
  as a function of the initial temperature $\Delta T(T_{\rm initial})$.
  It is predicted that the formation of domain structures under uniaxial tensile stress
  degrades the effectiveness of the elastocaloric effect.
}


\maketitle

\section{Introduction \label{introduction}}
Solid-state caloric effects provide a promising approach to future refrigeration technologies.
The electrocaloric~\cite{ECERochelle,ReviewECEScott,1956ECE,1961ECE},
magnetocaloric~\cite{RecentMagneto,FirstMagneto},
and barocaloric effects~\cite{BaroNiMnIn,BarocaloricThermalExpansionPhase}
produce a temperature change due to entropic changes induced by the application of
an electric field, magnetic field, and pressure, respectively.
The application of uniaxial stress to a ferroelectric material affects the spontaneous
polarization and produces an adiabatic temperature change. This is called the
elastocaloric
effect\cite{ElastocaloricMagnetocaloricShapeMemory,GiantElastocaloricBaSr,MulticaloricUSF,ElastoNiTiWires,ElastoMartensiticTransition}.

Here, a first-principles effective Hamiltonian model implemented
within a molecular dynamics (MD) framework is used to predict the elastocaloric response
of PbTiO$_{3}$.
Following the work of Lisenkov \textit{et al}.\cite{MulticaloricUSF},
the elastocaloric response of PbTiO${_3}$ is examined for tensile uniaxial loads ranging
from 0 to $-2.0$~GPa and temperatures ranging from
300 to 1000~K.
The results of this study will be compared with those reported in the literature.

\section{Methods \label{methods}}
The effective Hamiltonian used is
\begin{multline}
  \label{eq:Effective:Hamiltonian}
  H^{\rm eff}
  = \frac{M^*_{\rm dipole}}{2} \sum_{\bm{R},\alpha}\dot{u}_\alpha^2(\bm{R})
  + \frac{M^*_{\rm acoustic}}{2}\sum_{\bm{R},\alpha}\dot{w}_\alpha^2(\bm{R})\\
  + V^{\rm self}(\{\bm{u}\})+V^{\rm dpl}(\{\bm{u}\})+V^{\rm short}(\{\bm{u}\})\\
  + V^{\rm elas,\,homo}(\eta_1,\dots\!,\eta_6)+V^{\rm elas,\,inho}(\{\bm{w}\})\\
  + V^{\rm coup,\,homo}(\{\bm{u}\}, \eta_1,\cdots\!,\eta_6)+V^{\rm coup,\,inho}(\{\bm{u}\}, \{\bm{w}\}).
\end{multline}
Here, the collective atomic motion is coarse-grained by the local soft mode
vectors $\bm{u}(\bm{R})$ and
local acoustic displacement vectors $\bm{w}(\bm{R})$
of each unit cell at $\bm{R}$ in a simulation supercell.
$\eta_1,\dots,\eta_6$ are the six components of homogeneous strain in Voigt notation.
$\frac{M^*_{\rm dipole}}{2} \sum_{\bm{R},\alpha}\dot{u}_\alpha^2(\bm{R})$ and
$\frac{M^*_{\rm acoustic}}{2}\sum_{\bm{R},\alpha}\dot{w}_\alpha^2(\bm{R})$ are the
kinetic energies possessed by the local soft modes and
local acoustic displacement vectors along with their effective masses of $M^*_{\rm dipole}$ and $M^*_{\rm acoustic}$,
$V^{\rm self}(\{\bm{u}\})$ is the local-mode self-energy,
$V^{\rm dpl}(\{\bm{u}\})$ is the long-range dipole-dipole interaction,
$V^{\rm short}(\{\bm{u}\})$ is the short-range interaction between local soft modes,
$V^{\rm elas,\,homo}(\eta_1,\dots,\eta_6)$ is the elastic energy from homogeneous strains,
$V^{\rm elas,\,inho}(\{\bm{w}\})$ is the elastic energy from inhomogeneous strains,
$V^{\rm coup,\,homo}(\{\bm{u}\}, \eta_1,\dots,\eta_6)$ is the coupling between the local soft modes and the homogeneous strain, and
$V^{\rm coup,\,inho}(\{\bm{u}\}, \{\bm{w}\})$ is the coupling between the soft modes and the inhomogeneous strains.
Details of this Hamiltonian are explained in Refs.~\citen{SmithVanderbilt1994,ZhongVanderbiltRabe1995,feramderive}.
Additionally, to investigate the effects from stress,
we use the enthalpy $\mathcal{H}=H^{\rm eff}+N a_0^3\,\bm{\sigma}\cdot \bm{\eta}$,
where $N=L_x\times L_y\times L_z$ is the supercell size and
$a_0$ is the unit cell length; therefore, $N a_0^3$ is the supercell volume
and $\bm{\sigma}$ is the six components of stress.
In this study, we apply uniaxial tensile stress to the system along the $z$-direction.
It is implemented in the MD framework, and the MD simulation program is called \texttt{feram}.
\texttt{feram} is distributed as free software under the conditions described in
the GNU General Public License from its website\cite{feram}.
Examples of the input files are packaged within the source code under
the \texttt{feram-0.22.05/src/28example-PbTiO3-elastocaloric-770K/} directory.
The model parameters for PbTiO$_{3}$ are
determined semi-empirically in a previous work\cite{Waghmare:R:1997PRB}
and adopted for \texttt{feram} in Ref.~\citen{LeadParameters}.

\begin{figure}
  \centering
  \includegraphics[width=0.95\columnwidth]{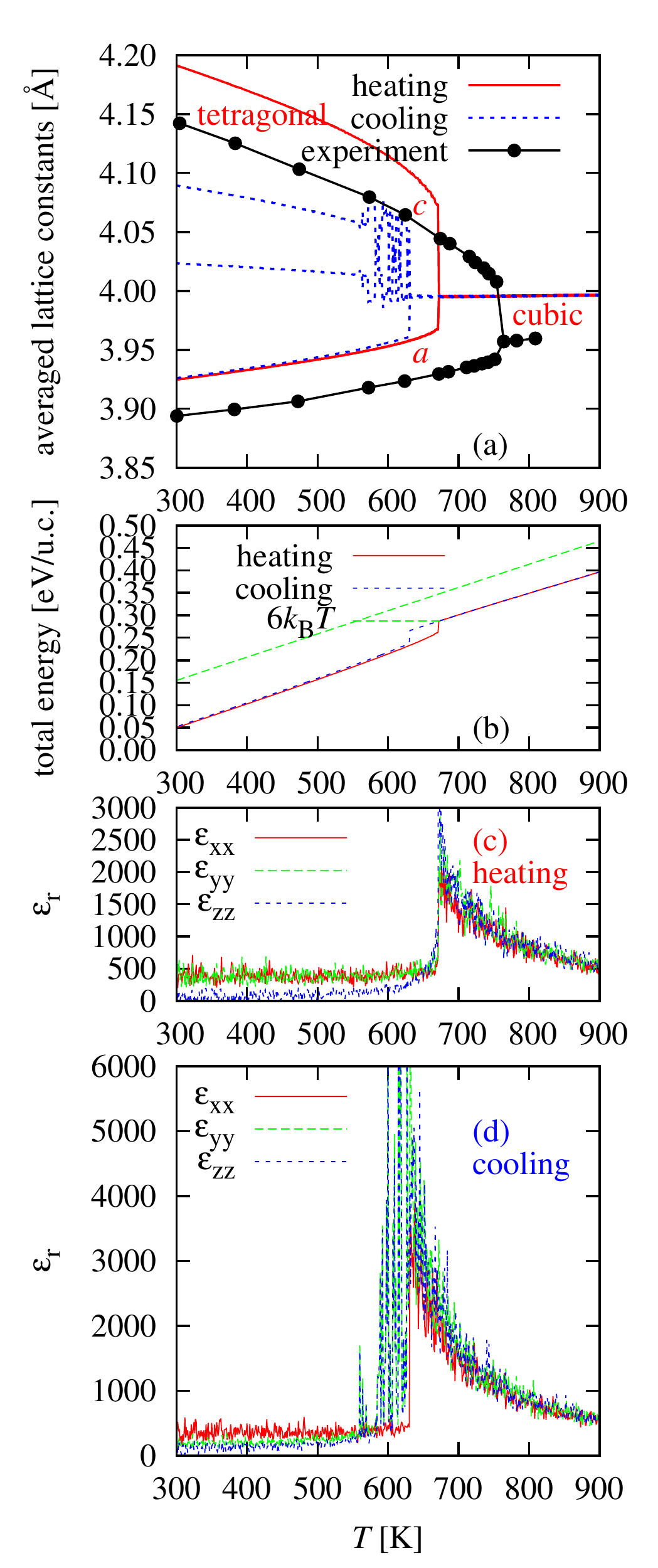}
  \caption{(Color online) Heating-up and cooling-down MD simulations of bulk PbTiO$_3$.
    (a) Averaged lattice constants.
    (b) Total energy per unit cell.
    Three components of relative dielectric constant are also plotted for
    (c) heating-up and
    (d) cooling-down simulations.
    Thermal hysteresis can be seen.
    In (a), experimentally observed
    lattice constants\cite{SHIRANE:H:S:PHYSICALREVIEW:80:p1105-1106:1950} are
    plotted with black solid lines.}
  \label{heating-cooling}
\end{figure}

Using the above parameters, the results of heating-up and cooling-down test MD simulations
for a supercell of $N = 16 \times 16 \times 16$
are shown in Fig.~\ref{heating-cooling}.
From the temperature $T$ dependences of the averaged lattice constants [shown in Fig.~\ref{heating-cooling}(a)],
a tetragonal-to-cubic ferroelectric-to-paraelectric
phase transition is clearly observed upon heating-up to 672~K.
During the cooling-down simulation,
$90^\circ$ ferroelectric domains are formed at 630~K
and are frozen at a low temperature, as described in Ref.~\citen{LeadParameters}.
These two transition temperatures are largely dependent on supercell size
with a slight dependence on the initial random configurations of $\{\bm{u}\}$.
However, their average of 653~K
is in good agreement with
those obtained in earlier Monte Carlo\cite{Waghmare:R:1997PRB} and MD\cite{LeadParameters} simulations,
in which the same set of parameters were used.
This is lower than the experimental value $T_{\rm C}=763$~K\cite{SHIRANE:H:S:PHYSICALREVIEW:80:p1105-1106:1950}.
This disagreement between simulations and experiments is unavoidable,
owing to the errors in the total energy of the first-principles calculations,
which is around 10~meV per unit cell.
In Fig.~\ref{heating-cooling}(b),
the temperature dependence of the total energy per unit cell is
plotted for heating-up and cooling-down test simulations.
At the transition temperatures, we observe jumps in the total energy, i.e., the latent heat.
In Figs.~\ref{heating-cooling}(c) and \ref{heating-cooling}(d),
the relative dielectric constant tensor computed from the fluctuations of the dipoles is plotted.
This is defined as
\begin{equation}
  \epsilon_{\alpha\beta} =
  \frac{1}{V \epsilon_0 k_{\rm B} T}
  [\langle p_\alpha p_\beta \rangle - \langle p_\alpha \rangle \langle p_\beta \rangle],
  \label{eq:epsilon}
\end{equation}
where $V$ is the volume of the supercell,
$\epsilon_0$ is the absolute dielectric constant of vacuum,
$k_{\rm B}$ is the Boltzmann constant,
$p_\alpha$ is the $\alpha (= x,y,z)$ component of the total electric dipole moment
in the supercell, $\bm{p} = Z^*\sum_{\bm{R}}\bm{u}(\bm{R})$,
$Z^*$ is the Born effective charge associated with a soft mode vector,
and the angle brackets $\langle\rangle$ denote the statistical time average\cite{PhysRevB.75.014111}.
Divergence in the dielectric constants at the transition temperatures can be clearly observed,
although it is slightly underestimated compared with
the experimentally observed value\cite{Remeika197037}.

Temperature-independent elastic coefficients
($C_{11} = 302$~GPa, $C_{12} = 132$~GPa, $C_{44} = 351$~GPa) determined from the cubic structure are used,
although they slightly depend on temperature even with this Hamiltonian of Eq.~(\ref{eq:Effective:Hamiltonian})
because strains $\eta_1,\dots,\eta_6$ and dipoles $\{\bm{u}\}$ couple
through $V^{\rm coup,\,homo}(\{\bm{u}\}, \eta_1,\dots,\eta_6)$.

\begin{figure}
  \centering
  \includegraphics[width=0.6\columnwidth]{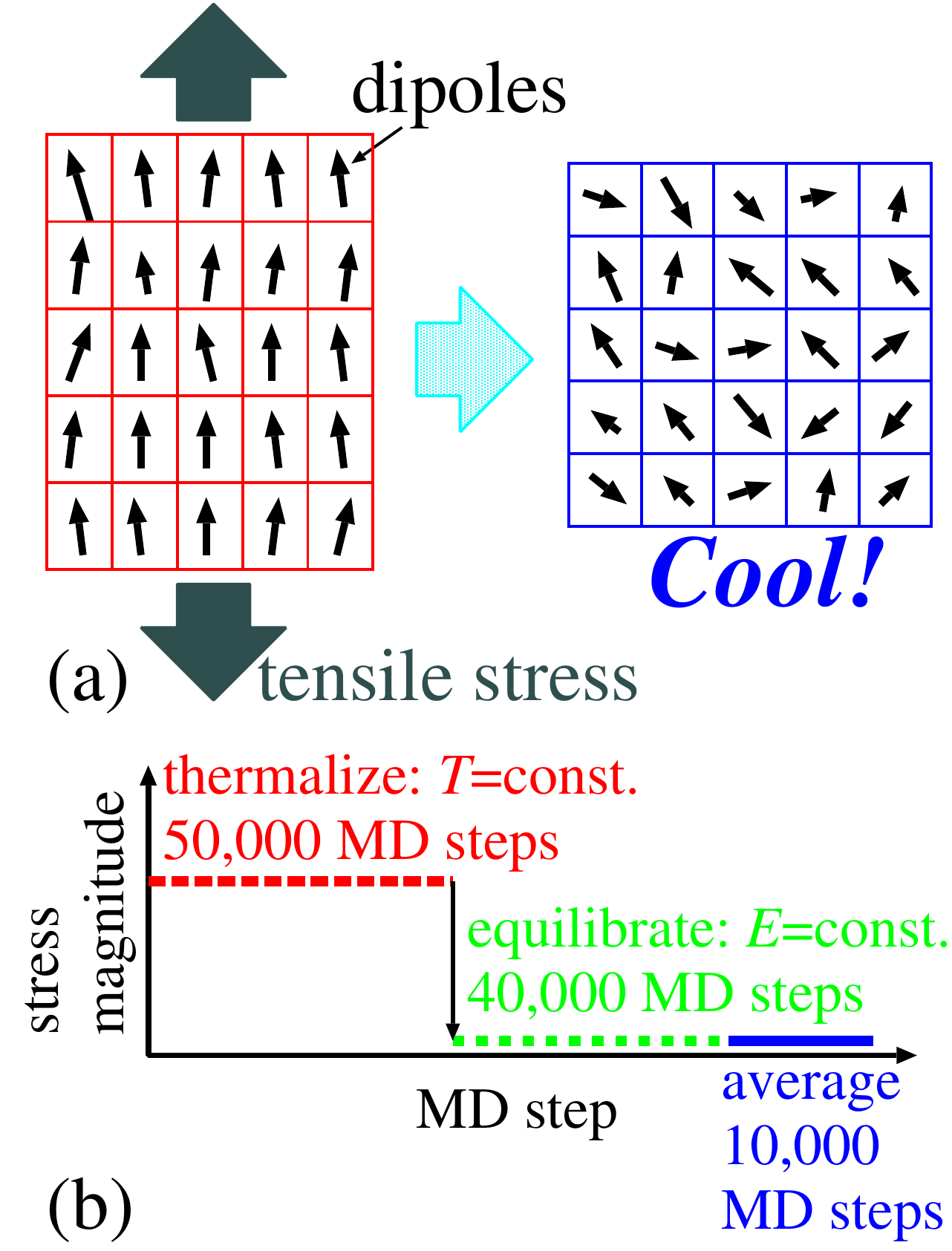}
  \caption{(Color online) (a) Schematic illustration of elastocaloric cooling.
    (b) Procedure of direct simulation of the elastocaloric effect.}
  \label{illustrations}
\end{figure}

As shown in Fig.~\ref{illustrations},
the simulation procedure for determining the elastocaloric response of PbTiO${_3}$
is similar to that used for the ``direct'' prediction of the electrocaloric effect
presented in Ref.~\citen{DirectIndirectBTO}.
A supercell size of $N = 64 \times 64 \times 64$ is used and is thermalized for
50,000 time steps in a canonical ensemble at the constant initial temperature $T_{\rm initial}$ and
constant applied stress.
A single-domain $+z$-polarized initial configuration for the first thermalization MD is generated randomly with
certain averages and deviations for $\{\bm{u}\}$:
$\langle u_x \rangle = \langle u_y \rangle = 0$, $\langle u_z \rangle = 0.33~{\rm \AA}$,
$\langle u_x^2 \rangle - \langle u_x \rangle^2 =
 \langle u_y^2 \rangle - \langle u_y \rangle^2 = (0.045~{\rm \AA})^2$, and
$\langle u_z^2 \rangle - \langle u_z \rangle^2 = (0.021~{\rm \AA})^2$.
Once thermalized, the system is switched
from being held at a constant temperature
to being isolated as a microcanonical ensemble.
The mechanical load is removed and the system is allowed
to equilibrate for 40,000 time steps.
Once equilibrated, the system's final temperature $T_{\rm final}$
is determined by averaging the acoustic and dipole kinetic energies for
10,000 time steps. The time step for this simulation is 2~fs.
One of the advantages of this ``direct'' prediction method
is that the temperature and external-field dependences of heat capacity and latent heat
are implicitly and automatically included in the simulations,
whereas in the ``indirect'' method,
an experimentally observed heat capacity must be used
in the entire temperature and external field ranges,
as described in Ref.~\citen{barrmatlet1}.
The temperature ranges from 300 to 1000~K, incremented with a step size of 1~K,
and the applied uniaxial stress ranges from 0 to $-2.0$~GPa, incremented
with a step size of $-0.2$~GPa.

\section{Results and Discussion\label{Results}}
\begin{figure}
  \centering
  \includegraphics[width=1.05\columnwidth]{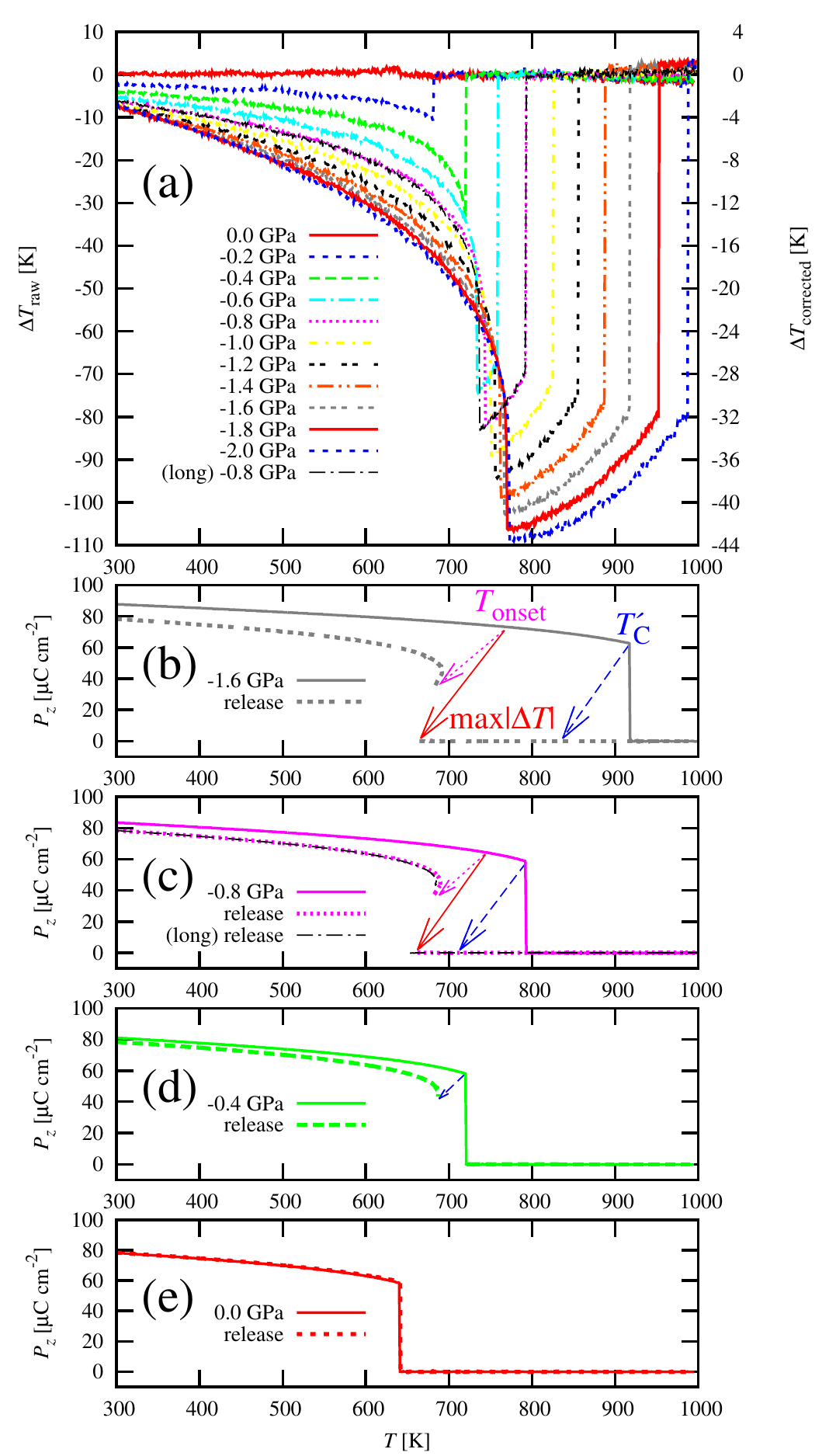}
  \caption{(Color online) (a) Simulated elastocaloric effect ($\Delta T$) in PbTiO$_3$ as a function of
    initial temperature ($T_{\rm initial}$).
    The applied uniaxial stress ranges from 0 to $-2.0$~GPa.
    $\Delta T$ is scaled from $\Delta T_{\rm raw}$ to $\Delta T_{\rm corrected}$
    by accounting for the reduced degrees of freedom, as discussed in Ref.~\citen{DirectIndirectBTO}.
    (b) Polarization along the $z$-axis both
    before [$P_z(T_{\rm initial})$] (gray solid line) and
    after  [$P_z(T_{\rm final})$] (gray dotted line) the release of load of $-1.6$~GPa.
    (c) $P_z(T_{\rm initial})$ (cyan solid line) and $P_z(T_{\rm final})$ (cyan dotted line) of load of $-0.8$~GPa.
    $P_z(T_{\rm final})$ after 990,000 MD time steps (thin black chain line).
    (d) $P_z(T_{\rm initial})$ (green solid line) and $P_z(T_{\rm final})$ (green dashed line) of load of $-0.4$~GPa.
    (d) $P_z(T_{\rm initial})$ (green solid line) and $P_z(T_{\rm final})$ (green dashed line) of zero load.
    In (b)--(d), transformations that give $T_{\rm onset}$, ${\rm max}|\Delta T|$, and $T'_{\rm C}$ are
    indicated by dotted magenta, solid red, and dashed blue arrows, respectively.}
  \label{ourPTO}
\end{figure}

The elastocaloric response $\Delta T_{\rm raw} = T_{\rm final} - T_{\rm initial}$ of PbTiO$_3$
is presented in Fig.~\ref{ourPTO}(a).
Scaling from $\Delta T_{\rm raw}$ to $\Delta T_{\rm corrected} = \frac{2}{5} \Delta T_{\rm raw}$
must be employed to account for the reduced degrees of freedom due to coarse graining,
as discussed in Ref.~\citen{DirectIndirectBTO}.
In Fig.~\ref{ourPTO}(b),
polarizations along the $z$-direction
before and after the release of the load of $\sigma_3=-1.6$~GPa are compared,
i.e., $P_z(T_{\rm initial},\,\sigma_3=-1.6~{\rm GPa})$
and  $P_z(T_{\rm final},\,\sigma_3=0)$ are compared, respectively.
We redefine the transition temperature under uniaxial stress as $T'_{\rm C}(\sigma_3)$.
For $\sigma_3=-1.6$~GPa, $T'_{\rm C}=917$~K.
Transformation from $T'_{\rm C}$ is indicated with a dashed blue arrow.
Above a certain temperature ($T_{\rm onset}$),
there is a temperature range $T_{\rm onset} < T_{\rm initial} \leq T'_{\rm C}$
in which one can obtain a large elastocaloric effect.
Transformation from $T_{\rm onset}$ is indicated by a dotted magenta arrow.
It can be seen that
below $T_{\rm onset}$ ($T_{\rm initial}\leq T_{\rm onset}$),
transformation from switching off the uniaxial tensile stress is
from an elongated ferroelectric polarized state
to a normal ferroelectric polarized state.
Between $T_{\rm onset}$ and $T'_{\rm C}$, i.e., $T_{\rm onset} < T_{\rm initial} \leq T'_{\rm C}$,
the transformation changes
from a stress-enhanced ferroelectric polarized state
to a paraelectric nonpolar state, resulting in a large elastocaloric response.
Just above $T_{\rm onset}$, a maximum $|\Delta T|$ is obtained
and its transformation is indicated by a solid red arrow in Fig.~\ref{ourPTO}(b).
Above $T'_{\rm C}$ ($T'_{\rm C}<T_{\rm initial}$), even under the uniaxial tensile stress load,
the system remains paraelectric and consequently $|\Delta T|=0$.

In Figs.~\ref{ourPTO}(c)--\ref{ourPTO}(e),
$P_z(T_{\rm initial}, \sigma_3\leq 0)$ and $P_z(T_{\rm final}, \sigma_3=0)$ are plotted also for
loads of $\sigma_3=-0.8$, $-0.4$, and $0.0$~GPa.
In Fig.~\ref{ourPTO}(c), it is observed that with a load of $-0.8$~GPa,
the effective temperature range $T_{\rm onset} < T_{\rm initial} \leq T'_{\rm C}$
becomes narrower than that of $-1.6$~GPa.
In Fig.~\ref{ourPTO}(d),
it can be seen that
the initial uniaxial tensile load of $-0.4$~GPa is not sufficiently large
to induce a ferroelectric-to-paraelectric transformation.
Therefore, we cannot define $T_{\rm onset}$ for loads of $0.0 < \sigma_3 < -0.4$~GPa.
In the case of zero load, in Fig.~\ref{ourPTO}(e),
the accuracy of our MD simulations ($\Delta T\equiv 0$) and
the simulated and underestimated phase transition temperature of $T_{\rm C}=640$~K under zero pressure are shown.
As anticipated, the greater the uniaxial loading, the greater the induced temperature
change $|\Delta T|$, and for a loading of $-2.0$~GPa, a temperature change of $-43$~K is
predicted.

\begin{figure}
  \centering
  \includegraphics[width=0.7\columnwidth]{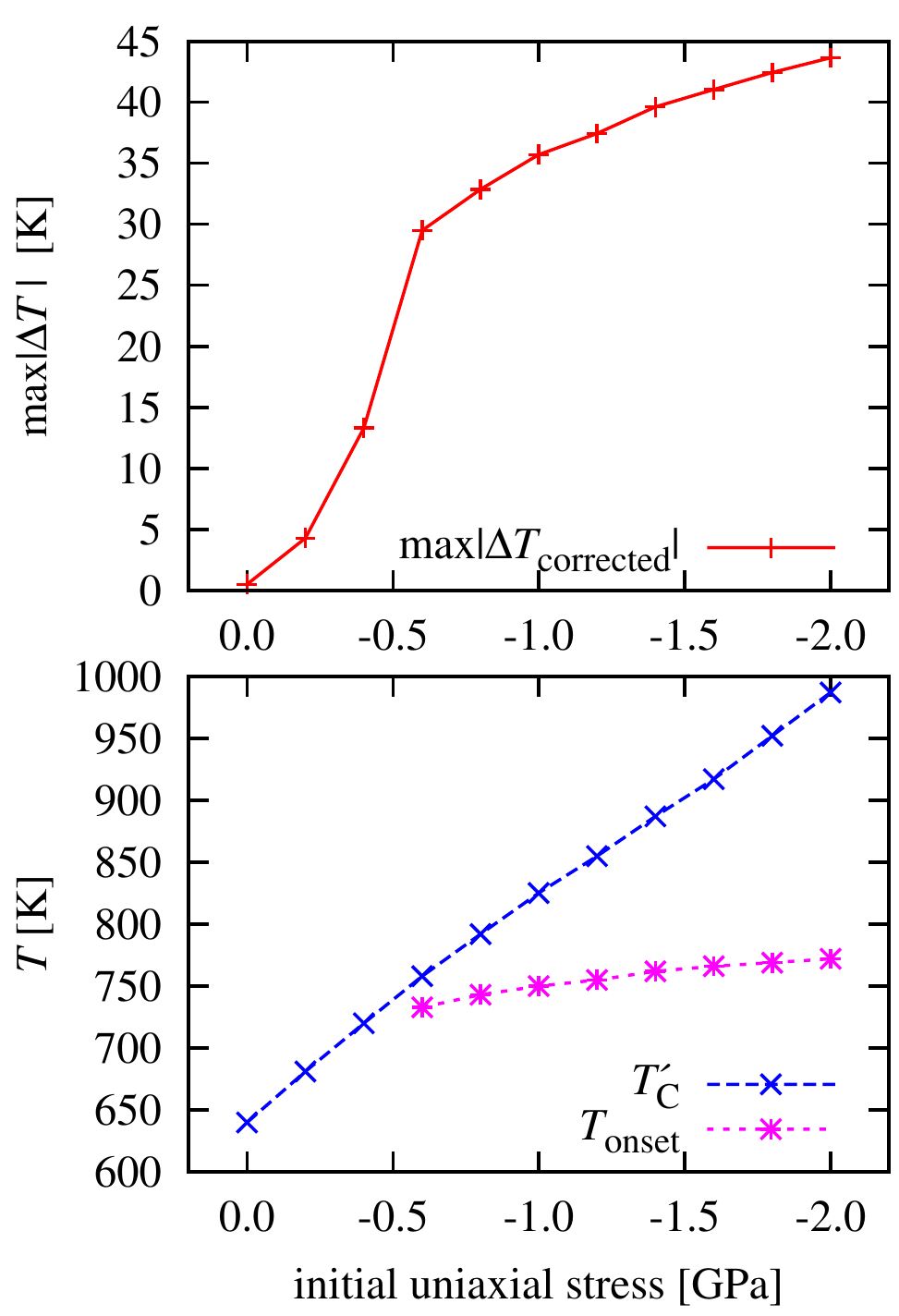}
  \caption{(Color online) Plots of ${\rm max}|\Delta T_{\rm corrected}|$, $T_{\rm onset}$, and $T'_{\rm C}$
    vs the different initially applied uniaxial tensile stresses.
    Data are connected with solid red, dotted magenta, and dashed blue lines, respectively.}
  \label{Comparison}
\end{figure}

In Fig.~\ref{Comparison},
plots show ${\rm max}|\Delta T_{\rm corrected}|$, and $T_{\rm onset}$ and $T'_{\rm C}$
under different applied loads.
It can be seen that $T'_{\rm C}$ linearly depends on applied load.
$T_{\rm onset}$ depends on applied load nearly linearly in $-0.6<\sigma_3<-2.0$, but less steeply
than $T'_{\rm C}$.

$T_{\rm onset}$ is also found to depend on the period of equilibration.
Between $T_{\rm C}$ and $T_{\rm onset}$ ($T_{\rm C}<T_{\rm initial}\leq T_{\rm onset}$),
when a uniaxial tensile stress is applied and then released,
the system stays in a ferroelectric state and does not transform into a paraelectric state.
In other words, the system {\it remembers} the strength of the stress applied.
This is confirmed with a longer equilibration of 990,000 time steps instead of the 40,000 shown in Fig.~\ref{illustrations}.
As indicated by black chain lines in Figs.~\ref{ourPTO}(a) and \ref{ourPTO}(c),
$T_{\rm onset}$ with longer equilibration becomes 736~K, whereas that of 40,000 was 744~K,
i.e., the system {\it forgets} the strength of  the stress applied.
Therefore, the stronger load and the shorter period of equilibration result in a higher $T_{\rm onset}$.

\begin{figure}
  \centering
  \includegraphics[width=0.99\columnwidth]{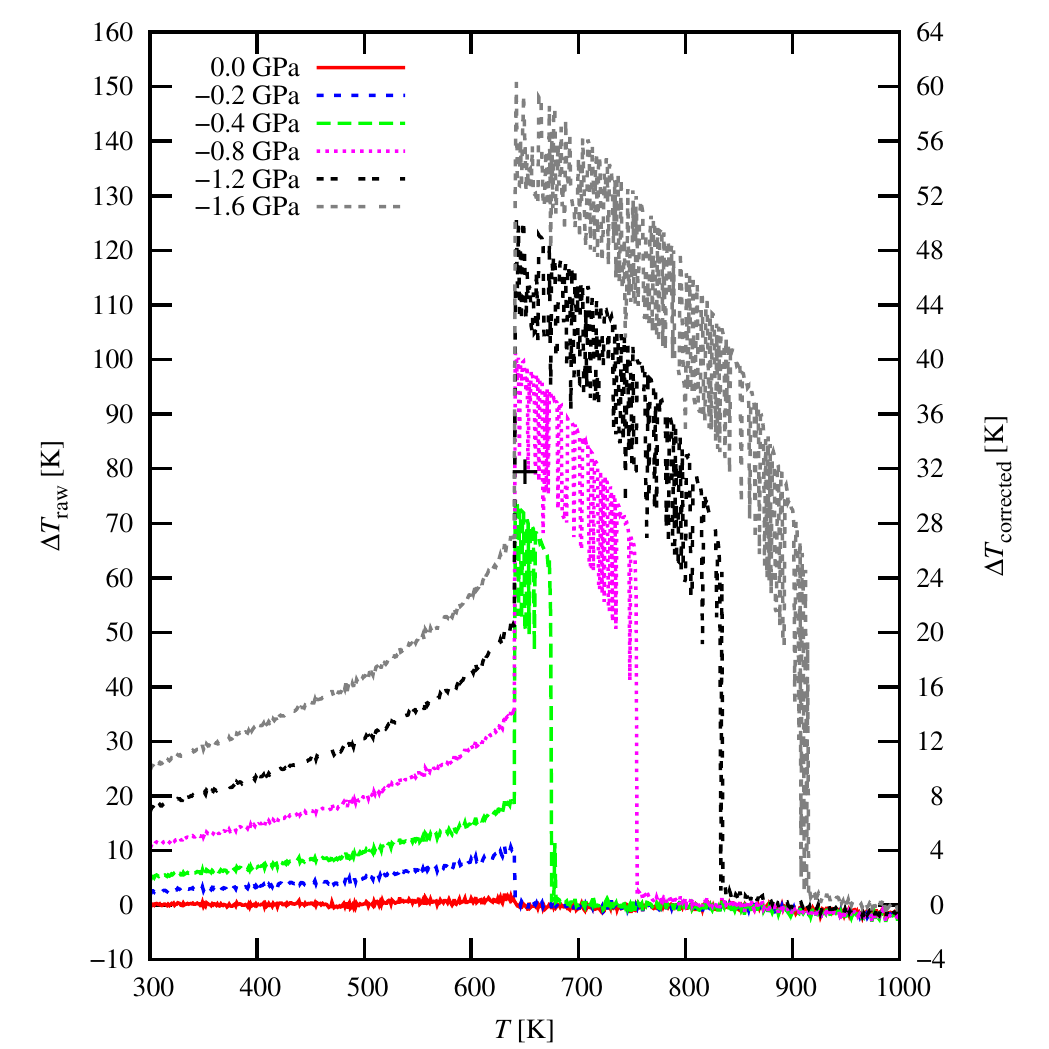}
  \caption{(Color online) Simulated switching-on elastocaloric effect with positive $\Delta T$
    in PbTiO$_3$ as functions of
    initial temperature, $T_{\rm initial}$.
    The switched-on uniaxial stress ranges from 0 to $-1.6$~GPa.
    $\Delta T$ is scaled from $\Delta T_{\rm raw}$ to $\Delta T_{\rm corrected}$.
    Zigzag structures in final temperatures are observed.
    A vertical cross section and a horizontal slice of a final state indicated with a ``+'' mark for a $-0.8$~GPa simulation are shown in
    Figs.~\ref{CrossSection}~and~\ref{slice}, respectively.}
  \label{OffOn}
\end{figure}

\begin{figure}
  \centering
  \includegraphics[width=0.99\columnwidth]{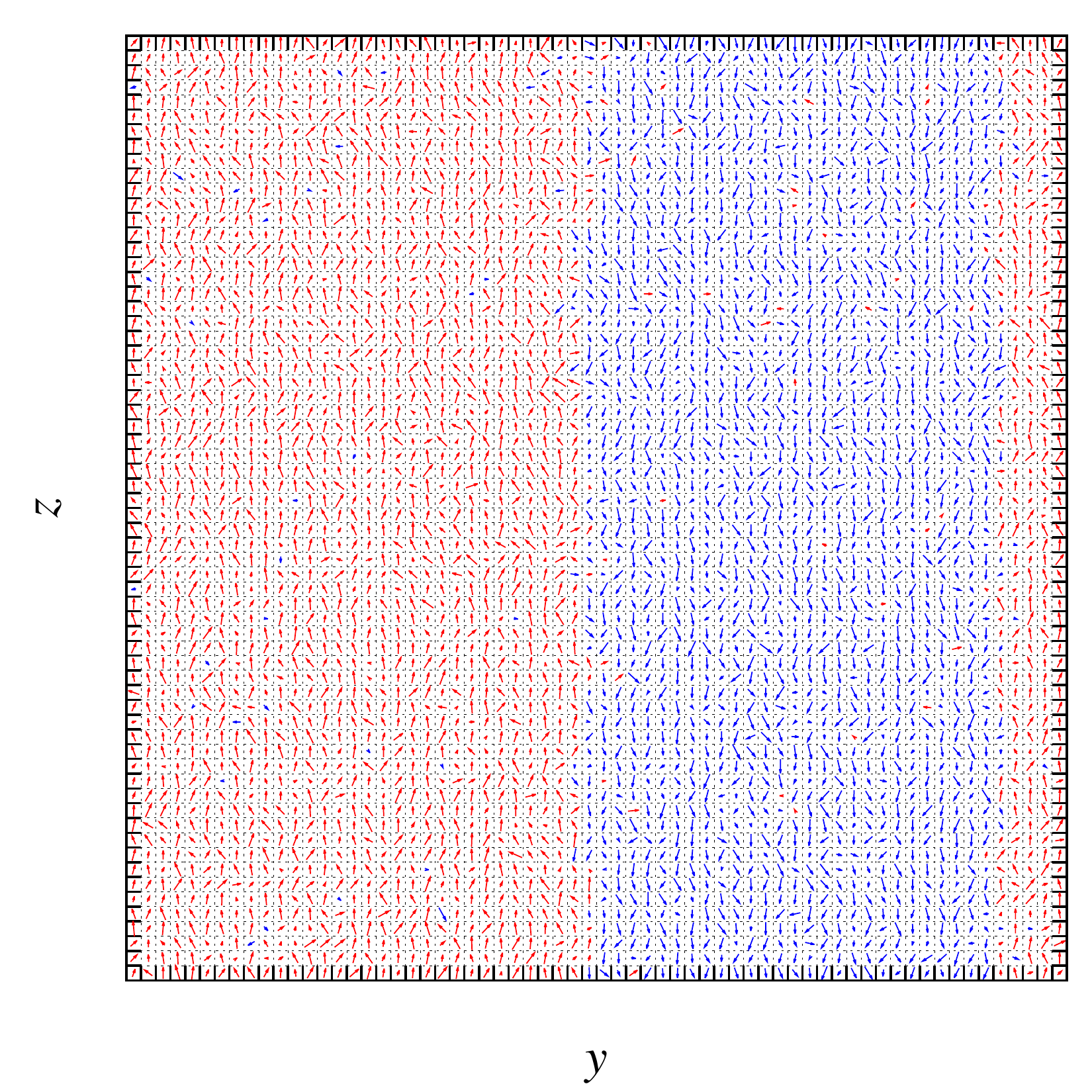}
  \caption{(Color online) Vertical cross section of
    a final state indicated with a ``+'' mark in Fig.~\ref{OffOn}.
    Dipole moments of each site are projected onto the $yz$-plane and indicated with arrows.
    The arrows are colored with red or blue if each dipole has $+z$ or $-z$ component, respectively.}
  \label{CrossSection}
\end{figure}

\begin{figure}
  \centering
  \includegraphics[width=0.86\columnwidth]{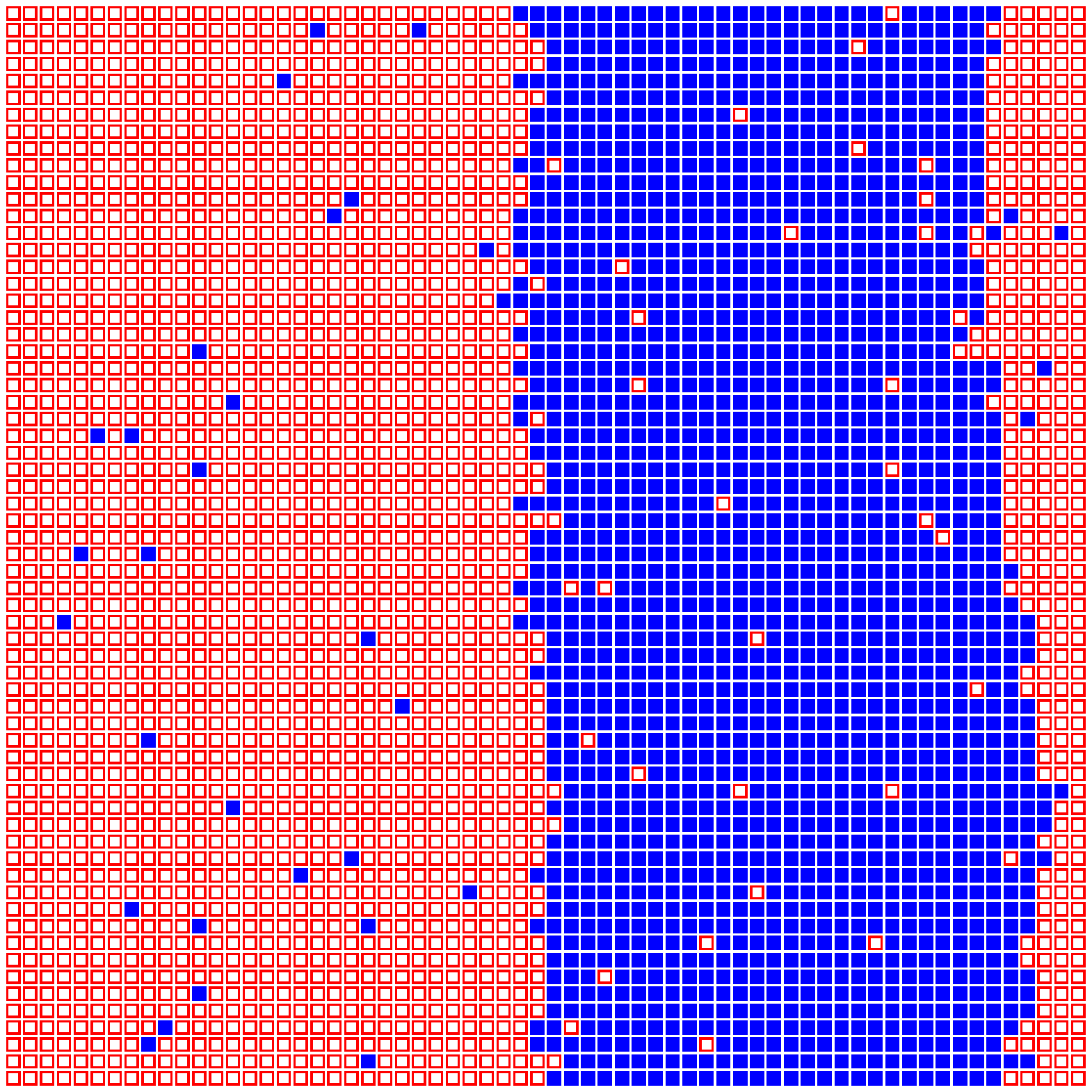}
  \caption{(Color online) Horizontal slice of
    a final state indicated with a ``+'' mark in Fig.~\ref{OffOn}.
    $+z$-polarized and $-z$-polarized sites are
    denoted by red $\square$ and blue $\blacksquare$, respectively.}
  \label{slice}
\end{figure}

In Contrast, in Fig.~\ref{OffOn}, we also perform ``heating'' simulations with switching-on of uniaxial stress
in which the system is firstly thermalized under zero stress and then
$\Delta T$ is measured under switched-on uniaxial tensile stresses.
Zigzag structures at the final temperatures are observed.
A vertical cross section and a horizontal slice of a final state indicated by a ``+'' mark in Fig.~\ref{OffOn} are shown in
Figs.~\ref{CrossSection}~and~\ref{slice}, respectively.
The supercell is divided into the $+z$ and $-z$ domains.
It can be understood that the zigzag structures arise from the existence and nonexistence of domain structures.
In the elastocaloric effect, domain structures may be formed more easily than in the electrocaloric effect,
because there is no significant $+z$- nor $-z$-direction in the uniaxial stress, but there is in the external electric field.
It is suggested that the formation of domain structures may cause some degradations in effectiveness in applications of the elastocaloric effect.
Note also that when comparing Figs.~\ref{ourPTO}(a)~and~\ref{OffOn},
the onset temperature is constant in switching-on ``heating'' simulations
because the simulations are started from zero stress.

The results presented here can be compared with those presented by
Lisenkov \textit{et al}.~in their Fig.~1(b)~\cite{MulticaloricUSF}.
Note that their $\Delta T$ is positive
because they switched on the uniaxial stress from zero stress.
For the same reason,
their onset initial temperature, which gives ${\rm max}|\Delta T|$, is always $T_{\rm C}$.
We have carried out {\it switching-off} time-dependent MD simulations and
found the applied-stress and equilibration-period dependences of $T_{\rm onset}$
because we consider that hysteretic behavior is important
for the cooling application of the elastocaloric effect.

Furthermore, whereas Lisenkov \textit{et al}. reported
a continuous linear increase in ${\rm max}|\Delta T|$ as stronger stresses are applied
and a maximum of approximately $+35$~K for a tensile load of $-2.0$~GPa,
our results of initial stress dependence of ${\rm max}|\Delta T|$ are not continuous at around $-0.5$~GPa.
Our MD simulation for a tensile load of $-2.0$~GPa results in ${\rm max}|\Delta T_{\rm corrected}|=|-43|$~K.

Finally, the shapes of the two $\Delta T$ vs $T$ plots differ in the high-temperature regime.
Both models have a ${\rm max}|\Delta T|$ that increases with loading, but the results here show a sharper
drop in $|\Delta T|$ at $T'_{\rm C}$, although this difference might be due to differences in the
implementations, not the underlying physics.
In Fig.~\ref{Comparison}, it
is observed that $T'_{\rm C}$ increases linearly with increased loading to a temperature of
1000~K, and presumably above, whereas Ref.~\citen{MulticaloricUSF} states that the
elastocaloric response disappears for temperatures above 890~K.

\section{Summary\label{Summary}}
Note that these simulations are very ideal and unrealistic ones.
For example, applying such huge tensile uniaxial stresses is difficult and
the phase transition of pure PbTiO$_3$ would cause
cracks in a crystal experimentally.
However, these ideal simulations suggest that, for example,
elastocaloric cooling has the largest effect
when a ferroelectric-to-paraelectric phase transition occurs.

In summary,
a first-principles effective Hamiltonian is used
in a molecular dynamics simulation to study the elastocaloric effect
in PbTiO$_3$.
The results show that for a modest loading of around $-0.5$~GPa, a thermal response
of around $-25$~K can be achieved, but for a large load of around $-2.0$~GPa, the
thermal response can be as large as $-44$~K.

The onset temperature $T_{\rm onset}$ and the termination temperature, $T'_{\rm C}$, are identified
as the temperatures bracketing the temperature range where the elastocaloric effect is greatest.
$T'_{\rm C}$ is found to scale linearly with initial load,
whereas $T_{\rm onset}$ has a less steep linearity.
Although increasing the initial stress widens the window of temperatures continuously,
the initial stress dependence of $\Delta T$ becomes smaller for stresses stronger than $-0.5$~GPa.

The formation of domain structures is observed in switching-on ``heating'' simulations and
it is suggested that the formation of domain structures
may cause some malfunctions in applications of the elastocaloric effect.

The results here are in qualitative agreement with those reported in
Ref.~\citen{MulticaloricUSF}, which were prepared using an effective
Hamiltonian in a Monte Carlo model; however, there are physically significant
differences including temperature and applied-stress dependences of $\Delta T$.
There is no easy explanation for these differences, and this requires future
investigation.

\section*{Acknowledgments}
The work of JAB and SPB was supported by the US
National Science Foundation (NSF) through grant No. DMR-1105641.
The NSF acknowledged for sponsoring JAB's travel to Tohoku University,
which was provided through grant No. DMR-1037898.
The work of TN was supported in part by JSPS KAKENHI Grant Number 25400314.
This work was also supported in part by the
Strategic Programs for Innovative Research (SPIRE), MEXT,
and the Computational Materials Science Initiative (CMSI), Japan.
The computational resources were provided by the Center for Computational Materials
Science, Institute for Materials Research (CCMS-IMR), Tohoku University.
We thank the staff at CCMS-IMR for their constant effort.
This research was also conducted using the Fujitsu PRIMEHPC FX10 System (Oakleaf-FX, Oakbridge-FX)
in the Information Technology Center, The University of Tokyo.

\bibliographystyle{prsty}
\bibliography{ElastoLeadv1}
\end{document}